\documentclass[aps,prd,preprint,showpacs,superscriptaddress,groupedaddress]{revtex4} 

\usepackage[latin1]{inputenc}
\usepackage{graphicx}
\usepackage{amssymb}
\usepackage{color}
\usepackage{float}
\usepackage{amsmath}
\usepackage{amsfonts}
\usepackage{wasysym}
\usepackage{dcolumn}
\usepackage{hyperref}
\hypersetup{
	colorlinks=true,       
	linkcolor=blue,          
	citecolor=blue,        
}
\usepackage{amsthm}
\usepackage{color}
\usepackage{amsmath}
\usepackage{mathrsfs}
\usepackage[dvipsnames]{xcolor}
\usepackage{xspace}

\def\3nab{\tilde{\nabla}}

\def\MT{M_{\rm T}}
\def\rs{r_{\rm s}}
\def\be {\begin{equation}}
\def\ee {\end{equation}}
\def\ba {\begin{eqnarray}}
\def\ea {\end{eqnarray}}

\newcommand{\bra}[1]{\left(#1\right)}

\newcommand{\brac}[1]{\left\{#1\right\}}

\newcommand{\sfr}[2]
{{\textstyle\frac{#1}{#2}}}

\newcommand{\lc}{\varepsilon}
\newcommand{\lb}{\{}
\newcommand{\rb}{\}}
\newcommand{\E}{{\mathcal E}}
\newcommand{\Q}{{\mathcal Q}}
\renewcommand{\H}{{\mathcal H}}
\newcommand{\barray}{\begin{array}}
\newcommand{\earray}{\end{array}}

\newcommand{\N}{N}

\newcommand{\hatn}{a}
\newcommand{\dotn}{\alpha}

\newcommand{\Es}{\mathscr{E}}
\newcommand{\Bs}{\mathscr{B}}

\newcommand \veps {\varepsilon}

\newcommand{\sss}[1][0.035cm]{\hspace*{#1}}
\newcommand{\A}{{\cal A}}

\newcommand{\udot}{{\mathcal A}}
\newcommand{\bea}{\begin{eqnarray}}
\newcommand{\eea}{\end{eqnarray}}
\newcommand{\f}[2]{\textstyle\frac{#1}{#2}}

\newcommand{\udota}{{\cal A}}
\newcommand{\nn}{{\nonumber}}
\renewcommand{\S}{_{\mathsf{S}}}
\newcommand{\T}{_{\mathsf{T}}}

\begin{document}
\title{Transient electromagnetic sources can detect solitary black holes in Milky Way galaxy}
\author{Susmita Jana}
 \email{susmitajana@iitb.ac.in}
 \affiliation{Department of Physics, Indian Institute of Technology Bombay, Mumbai 400076, India.}
\author{Rituparno Goswami}
 \email{goswami@ukzn.ac.za}
 \affiliation{Astrophysics Research Centre and School of Mathematics, Statistics and Computer Science, University of KwaZulu-Natal, Private Bag X54001, Durban 4000, South Africa.}
 \author{S. Shankaranarayanan}
 \email{shanki@iitb.ac.in}
 \affiliation{Department of Physics, Indian Institute of Technology Bombay, Mumbai 400076, India.}
  \author{Sunil D. Maharaj}
   \email{maharaj@ukzn.ac.za}
   \affiliation{Astrophysics Research Centre and School of Mathematics, Statistics and Computer Science, University of KwaZulu-Natal, Private Bag X54001, Durban 4000, South Africa.}
\begin{abstract}
The Milky Way galaxy is estimated to host up to a billion stellar-mass solitary black holes (BHs). The number and distribution of BH masses can provide crucial information about the processes involved in BH formation, the existence of primordial BHs, and the interpretation of gravitational wave (GW) signals detected in LIGO-VIRGO-KAGRA. {Sahu et al. recently confirmed one solitary stellar-mass BH in our galaxy using astrometric microlensing}. This work proposes a novel mechanism to identify such BH by analyzing the frequency and damping of the quasi-normal modes of GW generated from the interaction of the BH and EM wave originating from a transient electromagnetic (TEM) source. The incoming EM waves distort the curvature of a  BH, releasing GWs as it returns to a steady state. Using the covariant semi-tetrad formalism, we quantify the generated GWs via the \emph{Regge-Wheeler tensor} and relate the GW amplitude to the energy of the TEM. We demonstrate that isolated BHs at a distance of 50 pc from Earth can be detected by LIGO A+ and 100 pc by Cosmic Explorer/Einstein Telescope.     Additionally, we discuss the observational implications for orphan afterglows associated with GRBs, highlighting the potential for further discoveries.
\end{abstract}

\maketitle

\section{Introduction}
\label{sec:intro}
Transient astrophysical phenomena, typically high-energetic events, lead to rapid increase in luminosity over brief periods~\cite{2019-Murase.Bartos-AnnRevNuclPartPhys}. These events, originating from diverse astrophysical origins, exhibit distinct features~\cite{2019-Petroff-AARev,2004-Piran-RevModPhys}. For instance, Fast Radio Bursts (FRBs) emit energy between $10^{42}$ and $10^{44}$ ergs, over fractions of seconds in the MHz and GHz frequency bands~\cite{2023-Principe-AA}. Short duration gamma-ray bursts (GRBs) release substantial energy, around $10^{50}$ to $10^{52}$ ergs, within periods ranging from 10 milliseconds to several hours~\cite{2013-Berger-AnnRevAstAstPhy,2011-Wu.Wei.Etal-MNRS}. Similarly, supernovae events or long GRBs typically disperse energy ($10^{54}$ ergs) over several days to a month~\cite{2005-Jakobsson-A&A,2023-Srinivasaragavan-ApJL}.
{A GRB or similar transient events emit the amount of energy released is measured by the \emph{isotropic equivalent energy}, that lies between $10^{52} < E_{\rm iso} < 10^{54}$ erg for a long GRB emission~\cite{2018-A.Levan-GRB.Book}. Many predictions suggest existence of highly relativistic and beamed nature of long GRB emission~\cite{2012-Nemmen.etal-Science,2018-Lu.Kumar-MNRAS}. As per these predictions the true energy of the beamed GRB is $E_\gamma = E_{\rm iso} \left(1-\cos \theta_j \right)$, where $\theta_j$ is the $1/2$ of the opening angle of the GRB jet~\cite{2018-A.Levan-GRB.Book,2018-B.Zhang-grb.book}.  }

Traditionally, these events are observed in the electromagnetic (EM) window (for details, see Appendix.~\eqref{App-TEM}). However, with significant advancements in technology and response times, we can now detect these high-energy events not just through EM waves but also through gravitational waves (GWs)~\cite{2017-LIGOScientific-PRL}, neutrinos or cosmic rays~\cite{2019-Meszaros.etal-NPhys,2019-Murase.Bartos-ARNPS,2020-IceCube-ApJ}. The progress in detection capabilities underscores the rapid evolution of this field.

Despite the sensitivity of new-generation space-borne experiments and ground-based detectors to these events, the current detection rate stands at approximately one GRB or FRB per day~\cite{2009-Meegan-ApJ}. However, the limited field of view and the need for swift responses pose challenges in detecting transient events at high redshifts~\cite{Ghirlanda:2015aua,Zhang:2018csb,Fryer:2021yaq,Lan:2024ijv,Fialkov:2023ott}. Since FRBs and GRBs are understood to emit directional radiation, detection relies on the detectors' alignment with these transient events. Identifying transient candidates in synoptic sky surveys is challenging, particularly distinguishing them from low signal-to-noise ratio (S/N) data~\cite{Topinka:2015snq}. While machine learning algorithms aid detection, they introduce risks of false positives, such as instrumental glitches or contamination from non-interesting astrophysical transients~\cite{Mahabal-2019,Goode-2022}. 
{Compounding this issue is the significant difference between the theoretically predicted luminosity of the emitted energy $E_{\rm iso}$ from EM transients and the observed true energy $E_{\gamma}$ associated with the events~\cite{2012-Nemmen.etal-Science}}. This naturally prompts the question: Can we indirectly detect these events using alternative methods or windows?

This work shows that such transient EM (TEM) sources can produce a detectable amount of GWs while interacting with an intervening { isolated }black hole (BH). The geometry or curvature of the spherically symmetric static BH gets distorted by the incoming EM waves. The distorted BH releases GWs as it returns to a steady state~\cite{1970-Vishveshwara-Nature,1992-Chandrasekhar-BHBook,1999-Kokkotas-LivRev}. This process, referred to as \emph{ringdown} (the final stage of the merger of two BHs, where the newly formed BH settles into a stable state), is routinely seen in LIGO when coalescing BHs merge to form a larger BH~\cite{2023-GWTC3-PRX}. The characteristic frequency and damping of the ringdown, or quasi-normal modes (QNMs), depend on the parameters characterizing the BH and are independent of the initial configuration that triggered the vibrations~\cite{1992-Chandrasekhar-BHBook,1999-Kokkotas-LivRev}. In this work, we propose different ringdown signatures~\cite{1999-Kokkotas-LivRev,2002-Cardoso.Lemos-PLB} from a solitary BH.

{In generating gravitational waves (GWs) from electromagnetic (EM) waves, certain mechanisms involve direct EM-GW coupling. One such mechanism is the \emph{Gertsenstein-Zeldovich} (GZ) effect, in which EM waves (or GWs) can be converted into GWs (or EM waves) in the presence of a strong magnetic field~\cite{1962-Gertsenshtein-JETP,1974-Zeldovich-JETP}. In this process, the magnetic field acts as a catalyst, enabling resonance-based conversion between EM and GW modes, such that the input (EM or GW) and output (GW or EM) have the same frequency due to resonance conditions. The GZ effect has been useful for explaining various astrophysical phenomena, such as the production of fast radio bursts (FRBs) and electromagnetic signals from black hole ringdown events~\cite{1999-Marklund.Dunsby-APJ,2004-Clarkson.P.K.S-APJ,2022-Kushwaha.etal-MNRAS}. However, our analysis here diverges from the GZ effect. Instead, we explore how an intense transient EM wave distorts the spacetime curvature around a spherically symmetric black hole, inducing perturbations. As the black hole returns to equilibrium, this process results in the emission of gravitational waves. Interestingly, this mechanism does not require a magnetic field or any other external catalyst.} 

The Milky Way is estimated to host around $10^8$ solitary BHs with an average mass of approximately $14 M_{\odot}$, along with $\sim 10^8$ BHs in binary systems, boasting an average mass of $19 M_{\odot}$~\cite{2019-Olejak.Sobolewska-AstronAstrophys, 1993-Brown.Bethe-ApJ}. Accurately determining this number and distribution of BH masses can provide crucial information about the processes involved in BH formation, the possibility of the existence of primordial BHs, and interpreting GW signals detected in LIGO-VIRGO-KAGRA~\cite{2016-Sasaki.etal-PRL,2016-Bird.etal-PRL,2016-Hayasaki.etal-PASJ}. Detecting solitary BHs presents considerable challenges using current EM detectors. However, astrometric microlensing has made BH identification feasible~\cite{Kains:2018vnd,OGLE:2022gdj}. 
However, the key difference is that, unlike weak gravitating objects, the BH is active and converts the incoming EM waves to GWs. Also, in the proposed mechanism, the TEM source and observing detector need not be in the line of sight. The only requirement is that a solitary BH in the galaxy intervenes between the TEM and the GW detector. Since the GW emission from a distorted BH is isotropic, we can indirectly detect transient events like FRBs or short GRBs due to QNMs~\cite{1992-Chandrasekhar-BHBook,1999-Kokkotas-LivRev}.  
As we show, stellar-mass BHs offer a unique perspective on these TEM phenomena, observable in LIGO A+, III-generation GW detectors such as Cosmic Explorer and Einstein Telescope in the kHz frequency range~\cite{2021-Evans.Satya-CEReport,Kalogera:2021bya,2022-Cahillane.Mansell-Galaxies,Coccia:2023wag,Grado:2023pyb,Evans:2023euw}.

To quantify the effect, we need to use BH perturbation theory. Traditionally, BH perturbation formalism uses the \emph{Regge-Wheeler gauge} to compute QNMs~\cite{1999-Kokkotas-LivRev,Maggiore:2018sht}. Although the Regge-Wheeler gauge helps extract near-horizon physics, it is not suited for describing emitted GWs for asymptotic flat spacetimes and when the source of perturbation is a point particle~\cite{Maggiore:2018sht}. In such situations, TT gauge is a more preferred~\cite{2009-Sathyaprakash.Schutz-LRR}. Thus, one has to transform the Regge-Wheeler gauge solution to the TT gauge to read the waveform of the GW radiated at infinity due to the perturbed BH. 
Hence, we choose a covariant and gauge invariant 1 + 1 + 2 semi-tetrad perturbation formalism to study perturbations~\cite{1996-vanElst.Ellis-CQG,2003-Clarkson.Barrett-CQG,chris,2020-Hansraj.Sunil.Maharaj-GRG}. In this setup, any geometrical and thermodynamic quantities that vanish in the background are automatically gauge invariant by Stewart-Walker Lemma~\cite{SW}. Therefore, there is no need to switch between gauges as we evolve the metric perturbations from the BH to the asymptotic observer.
%

\section{The setup:} \label{sec:setup:}
\begin{figure}
\centering
\includegraphics[scale=.135]{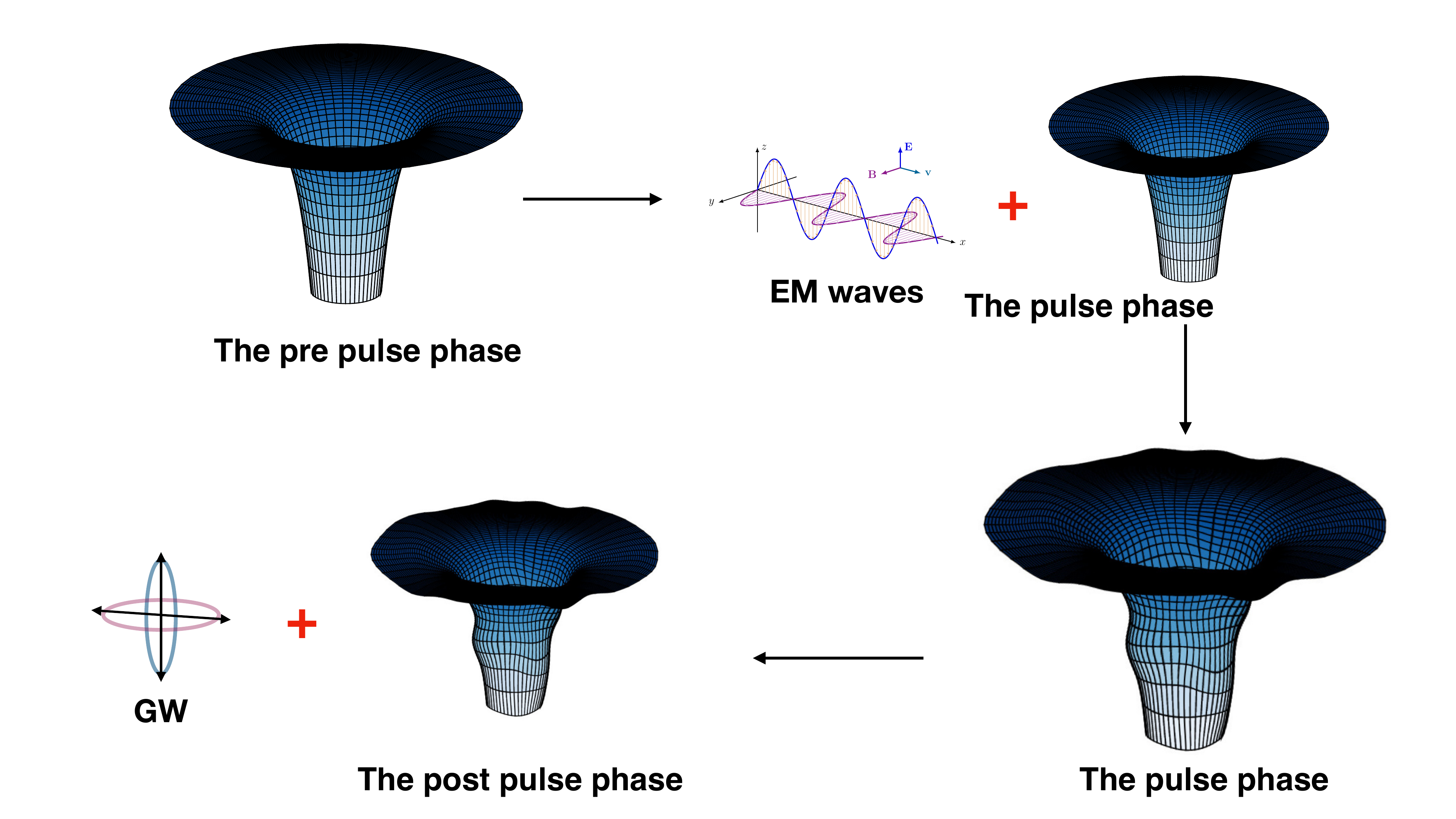}
\caption{Generation of GWs due to transient source.}
\label{fig:GZ-Grav}
\end{figure}

We consider the setup shown in Fig.~\ref{fig:GZ-Grav}. A high-energy 
TEM event distorts the spherically symmetric,
non-spinning Schwarzschild BH of mass $M$~\cite{AlmBir}. 
Specifically, the transient phenomenon generates a non-spherical EM pulse of finite duration. The distorted BH relaxes by emitting GWs.

%
%

To quantify this process, we employ the covariant $1+1+2$ semi tetrad formalism~\cite{1996-vanElst.Ellis-CQG,2003-Clarkson.Barrett-CQG,chris,2020-Hansraj.Sunil.Maharaj-GRG}. The observable quantities are evaluated w.r.t a comoving observer characterized by a 4-velocity $u^a$ satisfying $u^au_a=-1$. Additionally,  due to spherical symmetry, the spacetime features a preferred spacelike direction $n^a$ with $n^an_a=1$.
Thus, the 4-D spacetime metric can be decomposed as
\be
g_{ab}=-u_au_b+n_an_b+N_{ab} \, ,
\ee
where $N_{ab}$ is the {projection tensor} on the 2-D sheets orthogonal to $u^a$ and $n^a$. The directional derivative along $u^a$ is denoted by {\it dot derivative} `($\;\dot{}\;$)' and the fully projected directional derivative along $n^a$ is denoted by {\it hat derivative} `($\;\hat{}\;$)'.  The complete system is described by the geometrical variables of these two congruences (such as expansion, shear, or vorticity) along with the decomposed variables of the Weyl tensor and EM energy-momentum tensor.
 {For details, see Refs.~\cite{1996-vanElst.Ellis-CQG,2003-Clarkson.Barrett-CQG,chris,2020-Hansraj.Sunil.Maharaj-GRG} and Appendix.~\eqref{sec:CovariantFormalism}.

We divide the process into three --- \emph{the pre-pulse, the pulse, and the post-pulse} --- phases.
As shown in the top-left image of  Fig.~\eqref{fig:GZ-Grav}, the pre-pulse phase is represented by an isolated Schwarzschild BH. In the $1+1+2$ semi tetrad formalism, the vacuum spacetime is described by $\mathcal{D}_0=\{ \phi,\udot, \E\}$, where $\phi$ is the spatial expansion of the spacelike congruence ($n_a$), $\A$ is the observer's acceleration scalar, and $\E$ is the Weyl scalar extracted from the electric part of the Weyl tensor\cite{1996-vanElst.Ellis-CQG,2003-Clarkson.Barrett-CQG,chris,2020-Hansraj.Sunil.Maharaj-GRG}. 

As shown in the right-side images of Fig.~\eqref{fig:GZ-Grav}, during the pulse phase, the background spacetime gets distorted by an incoming EM wave. The $1+1+2$ description of the stress tensor of the test EM pulse is
%
\ba
T_{ab} &= &\mu u_{a}u_{b} + ph_{ab}+2 \Q \, e_{(a}u_{b)} +2 \Q_{(a}u_{b)}\nonumber\\
&& + \Pi (e_{a}e_{b}- \N_{ab}/2) + 2\Pi_{(a}e_{b)} + \Pi_{ab}\;.\label{EMTensor}
\ea 
A detailed description of the unperturbed BH spacetime and the input EM pulse is given in the Appendix.~\eqref{App-Generation-GW}.

As shown in the bottom-left image of  Fig.~\eqref{fig:GZ-Grav}, in the post-pulse phase, the distorted BH emits GWs until it reaches a steady state, a phenomenon known as "ring down"~\cite{1970-Vishveshwara-Nature,1992-Chandrasekhar-BHBook}. The characteristic frequency and damping of the ringdown, or QNMs, depend only on Schwarzschild mass $M$~\cite{1999-Kokkotas-LivRev}. To see transparently how the non-sphericity is radiated away via GWs, we construct the following dimensionless, covariant, gauge invariant, 
transverse trace-free tensor $M_{\{ab\}}$ \cite{chris, anne}
\be
M_{ab}=\phi \,r^2\,\zeta_{ab}/2 - r^2\,\E^{-1}\,\delta_{\lb a}W_{b\rb}/3\;.
\label{Mab}
\ee
$\zeta_{ab}$ represents distortion of the 2-D sheet and quantifies GW amplitude in free space. $M_{ab}$ is the {\em Regge-Wheeler tensor} and obeys the following closed wave equation for odd and even parity cases
\be
\ddot M_{\lb ab\rb}-\hat{\hat M}_{\lb ab\rb}-\A\,{\hat M}_{\lb ab\rb}
+\phi^2  M_{ab}-\delta^2 M_{ab} =0~.
\label{RMtensorwave}
\ee
Interestingly, the tensor $M_{ab}$ measures sheet deformation via the electric part of the Weyl scalar and the deformation tensor related to the preferred spacelike direction~\cite{2003-Clarkson.Barrett-CQG,2004-Clarkson.P.K.S-APJ}. In the pre-pulse phase $M_{ab}=0$, while $M_{ab} \neq 0$ in the post-pulse phase. Thus, $M_{ab}$ {\em contains the memory of the EM pulse~\cite{2023-Jana.Shanki-PRD}.} 

Spherical symmetry allows us to decompose the geometrical variables as an infinite sum of components relative to a basis of harmonic functions. This allows us to replace angular derivatives appearing in the equations with a harmonic coefficient. Following Ref.~\cite{chris1},
we introduce a set of dimensionless spherical harmonics $Q=Q^{(\ell,m)}$ ($m=-\ell,\cdots,\ell$), defined in the background, as eigenfunctions 
of the spherical Laplacian operator: $\delta^2 Q =-[\ell(\ell+1)/r^2] \,Q$,  $\hat{Q}= \dot{Q} = 0$. 
We expand the first-order
scalar ${\Psi}$ in terms of spherical harmonics
\be
{\Psi}=\sum_{\ell=0}^{\infty}\sum_{m=-\ell}^{m=\ell} {\Psi}\S^{(\ell,m)}
Q^{(\ell,m)} = {\Psi}\S \,Q,
\ee
where the sum over $\ell$ and $m$ is implicit in the last equality. The replacements that must be made for scalars when expanding the equations 
in spherical harmonics are
\be
\!\! {\Psi} ={\Psi}\S\, Q ~, \delta_a {\Psi} = r^{-1}{\Psi}\S \,Q_a~,   \veps_{ab}\delta^b {\Psi} =r^{-1}{\Psi}\S\, \bar Q_a \, ,        
 \ee
where the sums over $\ell$ and $m$ are implicit and $\bar Q_a $ is the odd parity vector harmonics. Note that the moment the EM pulse arrives, $\dot{W}_{a}$ is non-zero (even if other non-spherical $\mathcal{D}^{\rm geom}$ quantities are zero). 
 \be
\dot{W}_a = -(\phi/2) \left( \delta_{a} \Q \right) - (2/{3}) \delta_a \dot{\mu}\label{newEdot1} 
\ee
Since the background space-time is  Schwarzschild, we set $u^{a} \equiv (\, 1/\sqrt{F(r)},\, 0,\, 0,\, 0)$.
Using energy conservation, we get $\dot{\mu} = - \hat{\Q} - (\phi + 2\A) \Q $. Ignoring the radial dependence of the energy flux near the BH ($\hat{\Q} =0$) and integrating the resultant for a short time $\Delta t$, we get: 
\ba 
W_{a}= \frac{\phi - 4\A}{6(\phi + 2\A)} \delta_{a} (\Delta \mu)  \, ,
\label{Sol-Edot}
\ea 
where $\Delta \mu$ is the EM energy transferred to the BH by a transient astrophysical source \emph{flaring up} for a short time $\Delta t$. Decomposing
$\Delta \mu = \Delta{\mu}\S \,Q,$
$M_{ab} = {M}\T \,Q_{ab}$ ($Q_{ab} = r^2 \,\delta_{\lb a}\delta_{b \rb}  \,Q $ is the even parity tensor spherical harmonics) and substituting the above equation in Eq.~\eqref{Mab}, we have:
\ba 
{M}\T \,Q_{ab}&  = -\dfrac{r^2 (\phi - 4\A)}{18 \E (\phi + 2 \A)}( \Delta{\mu}\S) \delta_{\lb a}\delta_{b \rb} Q \label{Mab1}
\ea
Thus, EM pulse from a transient source with energy $\Delta {\mu}\S$ { near the BH} will induce the following GW amplitude:
\ba 
{M}\T \,  = \frac{r^3}{36 M } \left(1 + \frac{3M}{M-r}  \right) \Delta {\mu}\S \, . \label{Mab-Amplitude1}
\ea 
 
\noindent This is the key result of this work, regarding which we want to discuss the following points: 
First, ${M}\T$ derived above is for the fictitious observer near the BH horizon. We must transform the above quantity to the static (detector) frame to compare with the observations. Since, ${M}\T$ is gauge and frame invariant~\cite{2003-Clarkson.Barrett-CQG}, in asymptotically flat spacetime ($\E \sim 1/r^3, \phi \sim 1/r$), we have $M_{ab} \sim r \zeta_{ab}$. $\zeta_{ab}$ is related to plane GW in Minkowski spacetime via the relation: $
\zeta_{ab} \sim (1/2) \partial_{z} h_{ab}^{\rm TT} 
$~\cite{2004-Clarkson.P.K.S-APJ}. {As mentioned earlier, the covariant formalism does not require switching between gauges, and $M_{ab}$ provides a way to evolve metric perturbations from BH to the GW detector.}
Second, the outgoing GW amplitude depends on the EM energy density~\cite{2023-Jana.Shanki-PRD}. $\Delta \mu_S$ is large for energetic transient events, leading to an appreciable change in GW amplitude.  
We demonstrate that LIGO A+ and third-generation GW experiments can detect this. 

\begin{figure}
\centering
\includegraphics[scale=0.4]{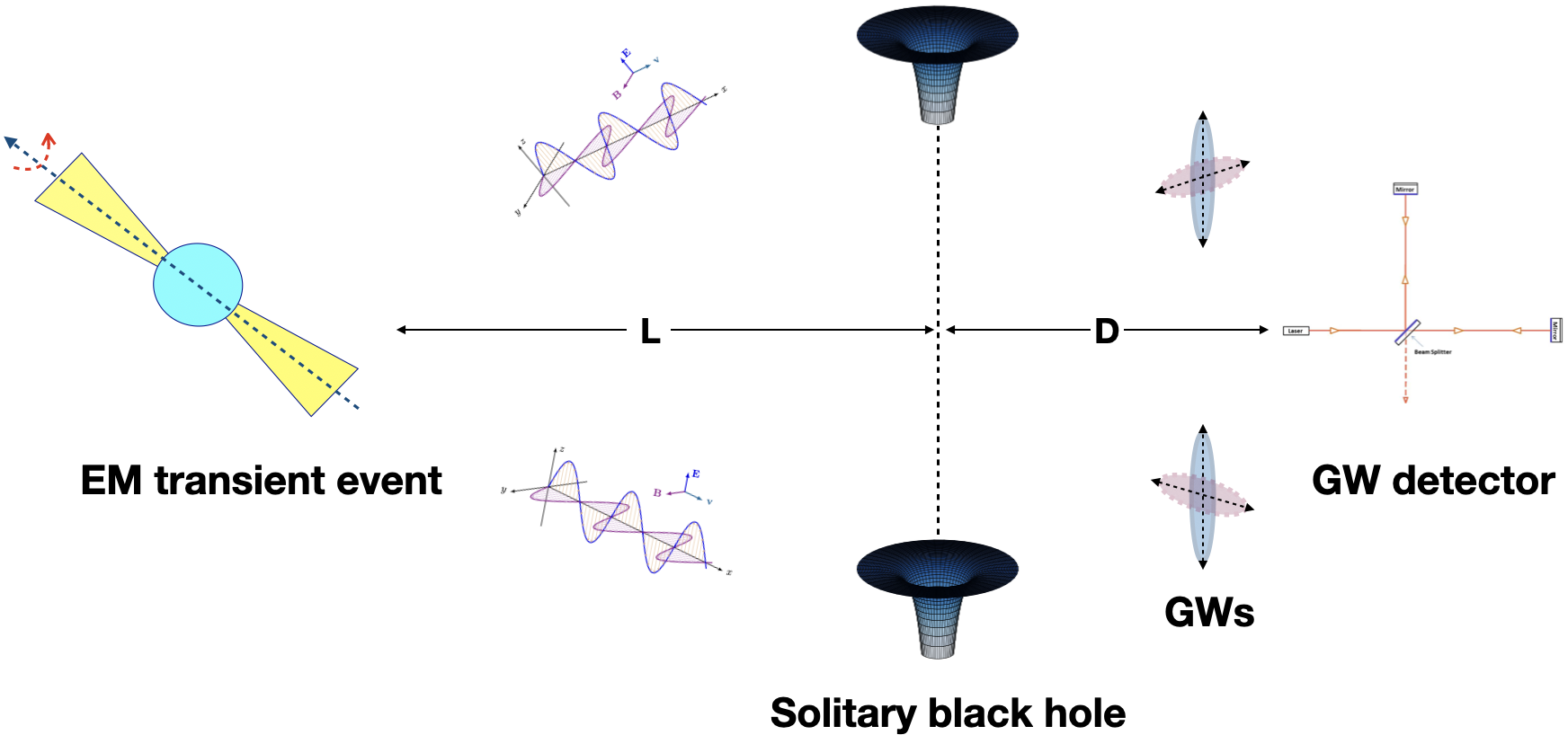}
\caption{Detection of GW signal due to intervening BH.}
\label{fig:Pulsar-BH-detector}
\end{figure}

\section{Observational implications} \label{sec:Observationaimplications}

{Fig.~\ref{fig:Pulsar-BH-detector} illustrates the scenario where $E_\gamma$  rapidly increases over a short duration ($\Delta t$) due to an energetic beamed TEM event. An energetic TEM event distorts the spherically symmetric BH at a distance of $L$ from the transient source. }
Subsequently, the distorted BH emits GWs until it reaches a steady state \cite{1970-Vishveshwara-Nature}. The generated GWs propagate to the detector at a distance $D$ from the BH. Since the ringdown modes exhibit characteristic frequencies determined solely by the BH parameters, their detection indirectly allows us to observe TEM events, such as FRBs and GRBs~\cite{Maggiore:2018sht}. Unlike EM counterparts, GW detectors do not need to be aligned along the line of sight of the EM transient; instead, the sole requirement is that the intervening solitary BH is along its line of sight. This relaxation enhances the detection of such transient events. To achieve this, we employ the following three steps: \\
1. \emph{GW amplitude ($\left. \MT  \right|_{\rm BH} $) near the BH horizon:}

{We begin by considering Eq.~\eqref{Mab-Amplitude1}. As detailed previously, the dimensionless variable $M_T$ and the original dimensionless variable $M_{ab}$ satisfy the \emph{Regge Wheeler} equation, describing the perturbation of a spherically symmetric BH~\cite{1992-Chandrasekhar-BHBook}. {In Eq.~\eqref{Mab-Amplitude1} $\Delta\mu\S$ represents the EM energy density of the TEM close to the BH horizon, falling on the BH and generating GW. It is 
expressed in terms of  $E_{\gamma}$ of the EM pulse~\cite{2016-NRAO-Book}}:
$\Delta\mu\S = \left(E_\gamma\right)/ \Delta V
$, where $\Delta V = \left(4\pi/3\right)  r^3$, where $r$ is distance from the BH center.}
Substituting this into Eq.~\eqref{Mab-Amplitude1}, the GW amplitude {in non-geometrised unit}, at a distance $r$ from the BH center is:
\ba 
\label{eq:MTatr0}
\MT = \frac{1}{4\pi} \, \frac{E_\gamma}{\mathcal{L}_0}\frac{c }{\rs} \left(1 + \frac{3}{1-\frac{2r}{\rs}} \right)
\ea
Here, $\mathcal{L}_{0} := c^{5}/G= 3.64 \times 10^{59}~{\rm erg/s}$ denotes the fundamental luminosity in gravitating systems~\cite{1985-Schutz-Book}. GW amplitude near BH, at 
a distance $r = 3\rs$ (corresponding to twice the radius of the photon sphere) is:
\ba 
\left. \MT  \right|_{\rm BH} = \frac{1}{10\pi} \frac{E_\gamma}{\mathcal{L}_0}\frac{c }{\rs}
  \, .
\label{eq:MTatr0-LEM-value}
\ea 
For a given TEM event, EM energy $E_\gamma$ is a fixed value. Thus, we see that 
$\MT$ depends on two factors --- the energy associated to the EM wave and the size or the mass of the BH $\rs$. The amplitude of the generated GW will be larger if $E_\gamma$ is higher and $r_s$ is smaller. For a stellar-mass BH of mass $M=10 M_{\odot}$ and $\rs = 2.9 \times 10^6$cm, 
we have $\left. \MT  \right|_{\rm BH} = 104.92 \times 
\left({E_\gamma}/{\mathcal{L}_{0}}\right)$.  
%
%

\noindent 2. \emph{GW Luminosity, frequency, and the damping factor:} GW energy radiated from the perturbed BH is~\cite{2004-Clarkson.P.K.S-APJ}:
\begin{align}
\frac{dE_{\rm GW}}{dt} = 
\, \mathcal{L}_{0} \times \left(  \left. M_{\rm T}\right|_{\rm BH} \right)^2 \, . \label{eq:Luminosity-GW2} 
\end{align}
Substituting Eq. \eqref{eq:MTatr0-LEM-value} in the above expression, we have:
\begin{align}
{\frac{dE_{\rm GW}}{dt} =
\frac{1}{100\pi^2} \frac{E_\gamma^2}{\mathcal{L}_0}\left(\frac{c }{\rs}\right)^2}
%
\label{eq:Luminosity-GW-Value} 
\end{align}
To determine the frequency of the outgoing GW, we consider the wave equation \eqref{RMtensorwave}, where $\delta^2 M_{ab}$ for the Schwarzschild background (in geometric units) is~\cite{2003-Clarkson.Barrett-CQG}:
\be
\delta^2 M_{ab} = \left(\phi^2 - 3\, \E - {l(l+1)}/{r^2} \right) \, M_{ab} \, .
\ee
Substituting the ansatz $M_{ab} = M_{\rm T}(r) \ e^{i\omega t} \ Q_{ab}$ in Eq.~\eqref{RMtensorwave}, leads to the following constraint at $r = 3 r_s$:
\ba 
\!\!\!\!\! -\omega^2 + 3 \, \E + {l(l+1)}/{r_{\rm s}^2} = 0 
&{\rm where} & \E = - \frac{2M}{(3 r_s)^3}
\label{eq:GW-freq-constraint}
\ea
From the above expression, we obtain the fundamental frequency for the lowest mode $l=2$~\footnote{In the geometrized units $t$ has the units of length; hence, $\omega$ is length-inverse. In non-geometrized units, $t$ ($\omega$) has units of time (inverse of time). In rewriting all the expressions in non-geometrized units, we have included the appropriate $c$ factor~\cite{1984-Wald-Book}.}:
\ba 
\label{eq:QNM-Freq}
\omega^2_{\rm Fund} = {3}/{r^{2}_{\rm s}} \, .
\ea
Interestingly, the above expression is consistent with the earlier results for Schwarzschild~\cite{1992-Chandrasekhar-BHBook,1999-Kokkotas-LivRev}. Appendix.~\eqref{sec:RWZ-comparison} compares the approach with the standard BH perturbation theory~\cite{Maggiore:2018sht} and shows that the two approaches give identical result.
{
The corresponding frequency $f_{\rm GW}$ in non-geometrized units is~\cite{1999-Kokkotas-LivRev}:
\begin{align}
\label{eq:omega-non-geometrised}
f_{\rm GW} 
= 9.327 \left(\frac{M_{\odot}}{M}\right) {\rm kHz}
\end{align}
Note that the QNMs of a perturbed BH with a mass of $10 M_{\odot}$ gives 
$f_{\rm GW} = 0.93~{\rm kHz}$. 
This frequency range($\mathcal{O}(1)~{\rm kHZ}$)} coincides with the high-sensitivity band targeted by upcoming gravitational wave detectors such as Cosmic Explorer (CE) and Einstein Telescope (ET)~\cite{2021-Evans.Satya-CEReport}. We obtained $\omega_{\rm Fund}$ for a fixed $r$, ignoring the effective potential. However, it is easy to see that the imaginary component responsible for mode decay exhibits a similar dependence on $r_s$~\cite{1970-Vishveshwara-Nature,1992-Chandrasekhar-BHBook,1999-Kokkotas-LivRev}.

\noindent 3. \emph{GW amplitude at the detector ($h_{\rm T}$):} 

In the first step, we evaluated the GW amplitude ($\left. \MT  \right|_{\rm BH}$) originating from the ringdown modes of the perturbed BH near its event horizon. Given that GWs propagate isotropically, we can determine the GW amplitude at the detector located on Earth. In the weak gravity regime,  using the relation between $M_{ab}$ and the transverse-traceless plane GW propagating along the $z$-axis in Cartesian coordinates~\cite{2004-Clarkson.P.K.S-APJ}, we have:
\begin{align}
M_{ab} \sim r \partial_{z} h_{ab}^{\rm (TT)} .\label{eq:Mab-hab}    
\end{align}
Using the plane-wave approximation, $h_{ab}^{\rm (TT)}= h_{\rm T} e^{i\omega(t-z)}$ in the above expression, the GW amplitude at the detector is:
\begin{align}
\left. h_{\rm T} \right|_{\earth} = 
c \left. \MT \right|_{\rm BH}/(D \omega) \, .
\label{eq:MT-hT}    
\end{align}
where $D$ is the distance from the BH to the detector and $\omega$ is the GW frequency. Substituting Eq.~\eqref{eq:MTatr0-LEM-value} in the above expression and setting $\omega = 2\pi f_{\rm GW}$ of the QNM frequency of a stellar-mass BH, we get: 
\begin{align}
{\left. h_{\rm T} \right|_{\earth} = 
 \frac{1}{10\pi}~\frac{c}{D \left(2\pi f_{\rm GW}\right)}
 \frac{E_\gamma}{\mathcal{L}_0}\frac{c }{\rs}}
\label{eq:MT-hT3}
\end{align}
For $M =10 M_{\odot}$, substituting $f_{\rm GW}$ from Eq.~\eqref{eq:omega-non-geometrised}, we get: 
\begin{align}
\left. h_{\rm T} \right|_{\earth} = 5.48 \times {10^{-10}} 
\left(\frac{D}{1~{\rm pc}}\right)^{-1} \,\left (\frac{E_\gamma}{\mathcal{L}_{0}} \right) \, .
\label{eq:MT-hT4}
\end{align}
This is the GW amplitude generated by a transient EM source, with a duration of approximately one second, detectable at a GW detector at a distance $D$ on Earth. Assuming the solitary BH is located at a distance of $100~{\rm pc}$ from Earth, we obtain:
\begin{align}
\left. h_{\rm T} \right|_{\earth} = 5.48 \times {10^{-12}} 
\,\left ( {E_\gamma}/{\mathcal{L}_{0}} \right) \, .
\label{eq:MT-hT5}
\end{align}
For the generated GW to be detectable by LIGO A+, CE or ET, the GW amplitude must be $h_{\rm T} \sim 10^{-24}$ or higher~\cite{Coccia:2023wag,Grado:2023pyb,Evans:2023euw,Kalogera:2021bya}. This implies $E_\gamma = 10^{47} ~ {\rm erg}$ and above can lead to $h_{\rm T} \left. \right|_{\earth}  \geq 1.51 \times 10^{-24}$.
For the generated GW to be detectable by CE or ET, the GW amplitude must be $h_{\rm T} \sim 10^{-24}$ or higher~\cite{Coccia:2023wag,Grado:2023pyb,Evans:2023euw,Kalogera:2021bya}. If the solitary BH is a distance of $50~{\rm pc}$ and $\mathcal{L}_{\rm EM} \sim 10^{47}~{\rm erg/s}$, it can lead to $h_{\rm T} \left. \right|_{\earth}  = 3.01 \times 10^{-24}$, which is line with the sensitivity of LIGO $A_+$~\cite{2022-Cahillane.Mansell-Galaxies}. Given that FRBs and GRBs exhibit energies $E_\gamma$ higher than $10^{47}~{\rm erg/s}$ and assuming an isotropic distribution of solitary stellar mass BHs within our galaxy, our model suggests that next-generation GW detectors like CE and ET  should observe a significant number of ringdown signals from solitary BHs in our galaxy. This would offer direct evidence for solitary stellar-mass BHs in our galaxy and indirect evidence for high-energy transient events.

The key assumption of this model is that energetic transient EM events, like FRBs and GRBs \cite{2019-Murase.Bartos-AnnRevNuclPartPhys,2020-Luo.Zhang-MNRS} distort intervening Schwarzschild BHs, generating detectable GWs in third-generation GW detectors. While the origins of FRBs/GRBs remain elusive, their observations do not show any apparent anisotropy and are consistent with an isotropic distribution of the arrival directions~\cite{2004-Piran-RevModPhys,2012-Nemmen.etal-Science,2018-Levan-Book,2018-Platts-PhysRep,2020-vonKienlin-ApJ,Rafiei-Ravandi:2021hbw,2021-CHIME-FRB-arXiv,2022-Kushwaha.Shanki-astro-ph.HE}. Our approach significantly enhances the detectability of such events, as it requires the solitary BH to be situated in the intervening region between the source and the GW detector rather than directly along the line of sight. {We can evaluate the efficiency of this process by taking the ratio of outgoing GW luminosity versus incoming EM wave luminosity  $\eta \sim \left( \mathcal{L}_{\rm GW} /\left(E_{\gamma}/\Delta t\right) \right) \sim 10^{-9} $, where $\Delta t$ is duration the EM wave stays near the BH. We assume $\Delta t \sim \mathcal{O}(1)~{\rm sec}$.}
  
One intriguing observational implication of our approach is the identification of orphan afterglows associated with GRBs. Currently, orphan afterglows are linked to transient events resembling the long-wavelength afterglow of a GRB but observed without the GRB trigger~\cite{2018-B.Zhang-grb.book}. In our approach, we estimate the luminosity of these incoming EM transients by utilizing the detected GW amplitude and frequency. This estimation process, which is detailed in equations \eqref{eq:MT-hT3} and \eqref{eq:QNM-Freq}, allows us to establish the energy function of these transients through GW detection. This may offer valuable insights into their origins and potential progenitors. Follow-up observations of orphan afterglows could establish a direct correspondence between the GW-detected signal and the orphan afterglow, further validating our approach.

To conclude, this novel mechanism enables the detection of solitary BHs in the Milky Way galaxy. Though our analysis is for a stellar-mass BH, the analysis is applicable for any non-rotating compact object, including some exotic compact objects~\cite{Cardoso:2019rvt,Volkel:2020lcy}.
{ Our analysis employs the \emph{covariant perturbation formalism}, which is specifically tailored to address perturbations in type II LRS spacetimes~\cite{2003-Clarkson.Barrett-CQG}. While this approach is readily adaptable to BHss such as Schwarzschild-de Sitter or Reissner-Nordström, it is not directly applicable to rotating BHs. In Recently, it has been shown using $1+1+2$ decomposition of the Kerr metric 
that rotating, axisymmetric BHs do not classify as LRS spacetimes~\cite{2021-Hansraj.SunilMaharaj-EPJC}. Consequently, we are currently extending our analysis to accommodate slowly rotating Kerr spacetimes within this framework. It is worth noting that the differences in QNM frequencies between Kerr and Schwarzschild BHs are modest~\cite{2009-Berti.etal-CQG}.
\begin{eqnarray}
& M \omega_{20} = 0.4437 - i \, 0.0739 \hat{b}^{0.3350} \quad \text{Kerr  BH} \nonumber \\
& M \omega = 0.3737 - i 0.0890 \quad \text{Schwarzschild BH}
\end{eqnarray}
where $\hat{b} = 1 - a/M$, $a/M \in [0,0.99]$, and $a$ is spin of the BH. The oscillation frequency and damping time of the Schwarzschild black hole set an upper bound for the fundamental QNM frequency for BHs in GR. Thus, this analysis offers a realistic threshold for detecting isolated BHs within our galaxy. While this mechanism can be extended to coalescing binary BHs, the GW signals generated during binary mergers are generally much stronger than those produced by this mechanism, making detection of this secondary effect unlikely and observationally insignificant.}

{
When considering interaction of EMs near BHs, the role of accretion disk on our proposed mechanism becomes relevant. Although a thorough investigation is required, preliminary estimates indicate that this effect is minimal. Since photons carry momentum and exert pressure, there is an upper limit to luminosity where gravitational forces can balance outward radiation pressure --- the Eddington luminosity: $ L_{\text{Eddington}} = 1.36 \times 10^{38} \, (M/M_\odot) \, \text{erg/s}$~\cite{2011-Abramowicz.Fragile-LRR}. For a stellar-mass black hole, this yields $ L_{\text{Eddington}} = 1.36 \times 10^{38} \, \text{erg/s}$. However, the transient electromagnetic events considered here exhibit luminosities well beyond this limit, suggesting that any influence from the accretion disk on our mechanism would be minimal.} 

The approach in this work has one distinct advantage over astrometric microlensing~\cite{OGLE:2022gdj} --- it can probe BHs at larger distances. In the case of microlensing, the detection is through the measurement of photon energy, which falls off as $1/D^2$. However, our approach uses GW amplitude, which falls as $1/D$. Hence, it can be a definite probe for detecting solitary BHs at further distances. This approach 
offers direct evidence for solitary stellar-mass BHs in our galaxy and indirect evidence for high-energy transient events.

\section*{Acknowledgements}

The authors thank P. Kushwaha, I. Chakraborty, S. Chakraborty, S. Kapadia, J. Khoury, S. Malik, S. Mandal, A. Miller, D. Radice, B. S. Sathyaprakash,  for discussions/comments on the earlier draft.
SJ thanks the IITB-IOE travel fund to the U. of KwaZulu-Natal where this work was initiated. RG thanks the South African National Research Foundation
(NRF) for support. SS thanks SERB-CRG for support.


\appendix
\section{Highly energetic transient electromagnetic events}
\label{App-TEM}
\begin{table}[!htb]
\label{Table:Transients-energy}
\begin{align}
\nonumber
\begin{array}{|l|c|c|c|l|}
\hline \text{Source} & \text{Rate density (Gpc$^{-3}$ Yr$^{-1}$)} & \text{Luminosity (erg/s)}   & \text{Duration (s)} \\
\hline \text { Long GRB } & 0.1-1 & 10^{51}-10^{52} & 10-100 \\
\text { Short GRB } & 10-100 & 10^{51}-10^{52} & 0.1-1 \\
\text { Low-luminosity GRB } & 100-1,000 & 10^{46}-10^{47} & 1,000-10,000  \\
\text { GRB afterglow } & & <10^{46}-10^{51} & >1-10,000  \\
\text{ FRB} & 339^{+1097}_{-330} & 10^{41}-10^{44} & \\
\hline
\end{array}
\end{align}
\caption{List of high energy electromagnetic transients. The rate density, Luminosity, and Duration are from Refs.~
}
\end{table}
As mentioned in the Sec.~\eqref{sec:intro}, energetic transient astrophysical phenomena
lead to rapid increase in luminosity over brief periods~\cite{2019-Murase.Bartos-AnnRevNuclPartPhys}. These events, originating from diverse astrophysical origins, exhibit distinct features~\cite{2019-Petroff-AARev,2004-Piran-RevModPhys}.
The table~\eqref{Table:Transients-energy} lists these high-energy astrophysical events observable in the electromagnetic spectrum, their corresponding rate density, Luminosity, and duration~\cite{2019-Murase.Bartos-AnnRevNuclPartPhys,2020-Luo.Zhang-MNRS}. While the origins of these events remain elusive, their observations do not show any obvious anisotropy and are consistent with an isotropic distribution of the arrival directions~\cite{2004-Piran-RevModPhys,2012-Nemmen.etal-Science,2018-Levan-Book,2018-Platts-PhysRep,2020-vonKienlin-ApJ,Rafiei-Ravandi:2021hbw,2021-CHIME-FRB-arXiv,2022-Kushwaha.Shanki-astro-ph.HE}.


\section{Semiterad covariant formalism}
\label{sec:CovariantFormalism} 

In this section, we briefly recapitulate the semi-tetrad formalism developed in  \cite{Covariant,chris1}, which enables us to study the problem geometrically. The key point of the semi-tetrad decompositions is that they are local decompositions defined on any open set $\mathcal{S}$. In the first step of this decomposition, the properties of spacetime are studied with respect to a real or fictitious observer whose velocity is along the tangent of a timelike congruence. Thereafter, if the spacetime has certain symmetries like local rotational symmetry, a preferred spatial direction exists. The spacetime is then further decomposed using this preferred spatial congruence. The field equations are then recast in terms of the geometrical variables related to these congruences and the curvature tensor of the spacetime (suitably decomposed using the congruences).

\subsection{Semitetrad 1+3 formalism}\label{A1}

$1+3$ covariant formalism is a well-known formalism widely used to study relativistic cosmology in general relativity (GR). The spacetime is locally sliced into the timelike direction $u^{a} \equiv dx^a/d\tau$ and spacelike hypersurface, which is orthogonal to $u^a$. Here, $\tau$ is the affine parameter.
In the 1+3 formalism \cite{Covariant}, the timelike unit vector ${u^{a}}$ ${\left(u^{a}\sss u_{a} = -1\right)}$ is used to split the spacetime locally in the form ${\mathcal{R} \otimes \mathcal{V}}$, where ${\mathcal{R}}$ is the timeline along ${u^{a}}$ and ${\mathcal{V}}$ is the 3-space perpendicular to ${u^{a}}$. Thus, the metric becomes
\begin{equation}
 g_{ab} = - u_{a}\sss u_{b}+ h_{ab},
\end{equation}
where ${h_{ab}}$ is the projection tensor used to project any vector or tensor on 3-space perpendicular to $u^a$. $h_{ab}$ becomes metric of the $3-$space iff there is no twist or vorticity in the $3-$space. The covariant time derivative along the observers' worldlines, denoted by `${\sss\sss^{\cdot}\sss\sss}$', is defined using the vector ${u^{a}}$, as
\begin{eqnarray}\label{dot}
\dot{Z}^{a ... b}{}_{c ... d} = u^{e}\sss\nabla_{e}\sss Z^{a ... b}{}_{c ... d},
\end{eqnarray} 
for any tensor ${Z^{a...b}{}_{c...d}}$. The fully orthogonally projected covariant spatial derivative, denoted by `\sss ${D}$\sss', is defined using the spatial projection tensor ${h_{ab}}$, as
\begin{eqnarray}\label{D}
D_{e}\sss Z^{a...b}{}_{c...d} = h^r{}_{e}\sss h^p{}_{c}\sss...\sss h^q{}_{d}\sss h^a{}_{f}\sss...\sss h^b{}_{g}\sss\nabla_{r}\sss Z^{f...g}{}_{p...q},
\end{eqnarray}
with total projection on all the free indices. The covariant derivative of the 4-velocity vector ${u^{a}}$ is decomposed irreducibly as follows
\begin{eqnarray}
\nabla_{a}\sss u_{b} &=& -u_{a}\sss A_{b} + \frac{1}{3}h_{ab}\sss\Theta + \sigma_{ab} + \epsilon_{abc}\sss \omega^{c},
\end{eqnarray}
where ${A_{b}}$ is the acceleration, ${\Theta}$ is the expansion of ${u_{a}}$, ${\sigma_{ab}}$ is the shear tensor, ${\omega^{a}}$ is the vorticity vector representing rotation and ${\epsilon_{abc}}$ is the effective volume element in the rest space of the comoving observer. The vorticity vector $\omega^q$ is related to vorticity tensor $\omega^{ab}$ as: $\omega^a \equiv (1/2)\, \epsilon^{abc} \, \omega_{bc}$.

Furthermore, the energy-momentum tensor of matter or fields propagating in the spacetime, w.r.t timelike vector ${u^{a}}$, is given by
\begin{eqnarray} \label{3.Tab}
T_{ab} &=& \mu\sss u_{a}\sss u_{b} + p\sss h_{ab} + q_{a}\sss u_{b} + u_{a}\sss q_{b} + \pi_{ab},
\end{eqnarray}
where ${\mu}$ is the effective energy density, ${p}$ is the isotropic pressure, ${q_{a}}$ is the 3-vector defining the heat flux and ${\pi_{ab}}$ is the anisotropic stress.
 Angle brackets denote orthogonal projections of vectors onto the three space as well as the projected, symmetric, and trace-free (PSTF) part of tensors.
\begin{eqnarray} 
v_{<a>}& = &h^{b}{}_{a}\sss\dot{V}_{b},\label{angbrac1}\\ 
Z_{<ab>}& =& \bra{h^{c}{}_{(a}\sss h^{d}{}_{b)} - \frac{1}{3}h_{ab}\sss h^{cd}}\sss Z_{cd}.\label{angbrac2}
\end{eqnarray}
The Weyl quantities are decomposed as
\begin{eqnarray}
 E_{ab} &=& C_{abcd}\sss u^{c}\sss u^{d} = E_{<ab>}, \label{E}\\
H_{ab} &=& \frac{1}{2} \varepsilon_{ade}\sss C^{de}{}_{bc}\sss u^{c} = H_{<ab>}. \label{H}
\end{eqnarray}

\subsection{Semitetrad 1+1+2 formalism}\label{A2}

In the 1+1+2 formalism \cite{chris1}, the 3-space ${\mathcal{V}}$ is now further split by introducing the unit vector ${e^{a}}$ orthogonal to ${u^{a}}$ ${\left(e^{a}\sss e_{a} = 1, u^{a}\sss e_{a} = 0\right)}$. The $3-$space now has two parts---one is the spacelike direction $e^a$, and the second is the $2-$space orthogonal to $e^a$ as well as $u^a$, which we refer to as the $2-sheets$. The 1+1+2 covariantly decomposed spacetime is given by 
\begin{equation}\label{2.Nab}
g_{ab} = -u_{a}\sss u_{b} + e_{a}\sss e_{b} + N_{ab},
\end{equation}
where ${N_{ab}}$ ${\left(e^{a}\sss N_{ab} = 0 = u^{a}\sss N_{ab}, N^{a}{}_{a} = 2\right)}$ projects vectors onto 2-spaces called `2-sheets', orthogonal to ${u^{a}}$ and ${e^{a}}$.  We introduce two new derivatives for any tensor ${\phi_{a...b}{}^{c...d}}$:
\begin{eqnarray}
\label{hatderiv}
\hat{\phi}_{a...b}{}^{c...d} &\equiv& e^{f}\sss D_{f}\sss \phi_{a...b}{}^{c...d}, \\
\label{deltaderiv}
\delta_{f}\phi_{a...b}{}^{c...d} &\equiv& N_{f}{}^{j} N_{a}{}^{l} ... N_{b}{}^{g} N_{h}{}^{c} ... N_{i}{}^{d}  D_{j}\phi_{l...g}{}^{h...i}.
\end{eqnarray}
 
The  1+3 kinematical and dynamical quantities and anisotropic fluid variables are split irreducibly as
\begin{eqnarray}
A^{a} &=& \mathcal{A}\sss e^{a} + \mathcal{A}^{a}, \\
\omega^{a} &=& \Omega\sss e^{a} + \Omega^{a}, \\
\sigma_{ab} &=& \Sigma\left(e_{a}\sss e_{b} - \frac{1}{2}\sss N_{ab}\right) + 2\sss\Sigma_{(a}\sss e_{b)} + \Sigma_{ab}, \\
\label{2.qa} q_{a} &=& \Q\sss e_{a} + \Q_{a}, \\
\label{2.pia} \pi_{ab} &=& \Pi\left(e_{a}\sss e_{b} - \frac{1}{2}\sss N_{ab}\right) + 2\sss\Pi_{(a}\sss e_{b)} + \Pi_{ab},\\
E_{ab} &=& \mathcal{E} \left(e_{a}\sss e_{b} - \frac{1}{2}\sss N_{ab}\right) + 2\sss\mathcal{E}_{(a}\sss e_{b)} + \mathcal{E}_{ab}, \\
H_{ab} &=& \mathcal{H}  \left(e_{a}\sss e_{b} - \frac{1}{2}\sss N_{ab}\right) + 2\sss\mathcal{H}_{(a}\sss e_{b)} + \mathcal{H}_{ab}\,.
\end{eqnarray}
 
The fully projected 3-derivative of ${e^{a}}$ is given by
\begin{eqnarray}
D_{a}\sss e_{b} &=& e_{a}\sss a_{b} + \frac{1}{2}\sss\phi\sss N_{ab} + \xi\sss\varepsilon_{ab} + \zeta_{ab},
\end{eqnarray}
where traveling along ${e^{a}}$, ${a_{a}}$ is the sheet acceleration, ${\phi}$ is the sheet expansion, ${\xi}$ is the vorticity of ${e^{a}}$ (the twisting of the sheet) and ${\zeta_{ab}}$ is the shear of ${e^{a}}$ (the distortion of the sheet). 

The 1+1+2 split of the full covariant derivatives of ${u^{a}}$ and ${e^{a}}$ are as follows
\begin{eqnarray} \label{2.delAUb}
\nabla_{a}\sss u_{b} &=& -u_{a}\left(\mathcal{A}\sss e_{b} + \mathcal{A}_{b}\right) + e_{a}\sss e_{b} \left(\frac{1}{3}\sss\Theta + \Sigma \right) 
 + e_{a}\left(\Sigma_{b} + \varepsilon_{bc}\sss\Omega^{c}\right) + \left(\Sigma_{a} - \varepsilon_{ac}\sss\Omega^{c}\right) e_{b}\nonumber\\
 &&+ N_{ab}\left(\frac{1}{3}\sss\Theta - \frac{1}{2}\sss\Sigma\right) + \Omega\sss\varepsilon_{ab} + \Sigma_{ab}, \\
\nabla_{a}\sss e_{b} &=& -\mathcal{A}\sss u_{a}\sss u_{b} - u_{a}\sss\alpha_{b} + \left(\frac{1}{3}\sss\Theta + \Sigma \right)e_{a}\sss u_{b}  + \left(\Sigma_{a} - \varepsilon_{ac}\sss\Omega^{c}\right)u_{b}  + e_{a}\sss a_{b}\nonumber\\
 &&+  \frac{1}{2}\sss\phi\sss N_{ab}+ \xi\sss\varepsilon_{ab} + \zeta_{ab}.		
\end{eqnarray}

We can now immediately see that the Ricci identities and the doubly contracted Bianchi identities, which specify the evolution of the complete system, can now be written as the time evolution and spatial propagation and spatial constraints of an irreducible set of geometrical and electromagnetic (EM) variables.
The irreducible set of geometric variables
\begin{eqnarray}
\label{Dgeom}
\mathcal{D}_{geom}
 = 
\{\Theta, \sss \mathcal{A}, \sss\Omega, \sss\Sigma, \sss\mathcal{E}, \sss\mathcal{H}, \sss\phi, \sss\xi, \sss\mathcal{A}_{a}, \sss\Omega_{a}, \sss\Sigma_{a}, 
\sss\alpha_{a}, \sss a_{a}, \sss\mathcal{E}_{a}, 
\sss\mathcal{H}_{a}, \sss\Sigma_{ab}, \sss\zeta_{ab}, 
\sss\mathcal{E}_{ab}, \sss\mathcal{H}_{ab}\}
\end{eqnarray}
together with the irreducible set of EM variables
\begin{eqnarray}
\label{Dtherm}
\mathcal{D}_{\rm EM}&=&\brac{\mu, \sss p, \sss \Q, \sss \Pi, \sss \Q_{a}, \sss \Pi_{a}, \sss \Pi_{ab}},
\end{eqnarray}
make up the key variables in the 1+1+2 formalism.


\section{Generation of GWs due to transient source: Details}
\label{App-Generation-GW}
%
%
As mentioned in the Sec.~\eqref{sec:setup:}, we divide the process into three --- \emph{the pre-pulse, the pulse, and the post-pulse} --- phases. The rest of this section contains details of the three phases.

\subsection{Pre-pulse phase}

The pre-pulse phase will be a spherically symmetric vacuum around an open set $\mathcal{S}$ containing $\gamma$. It hence will be locally equivalent to a part of the maximally extended Schwarzschild solution in $\mathcal{S}$. As the fictitious observer is outside the BH, we can always assume the existence of a timelike Killing vector $\xi^a$, for which the volume expansion ($\Theta$) and the shear scalar ($\Sigma$) vanish. By aligning the tangent of $\gamma$ to this Killing vector $\xi^a$ at every point in this pre-pulse phase, we can make the observer static; That is, the directional derivatives of all geometrical variables along the observer congruence (dot derivatives) vanish.
The only non-zero geometrical variables in the background spacetime are \cite{chris1} 
\begin{eqnarray}\label{D0}
\mathcal{D}_0=\{ \phi,\udot, \E\},
\end{eqnarray}
where $\phi$ is the spatial expansion of the spacelike congruence $n^a$, $\udot$ is the acceleration scalar for the observer, and $\E$ is the Weyl scalar extracted from the electric part of the Weyl tensor. These satisfy the following propagation (hat derivative  {along radial coordinate}) equations  
\begin{eqnarray}
\hat\phi=-\E-\phi^2/2 \;,\;\;\hat\A= -\bra{\A+\phi}\A \;,\label{Aphihat}
\end{eqnarray}
with the constraint $\E=-\A\phi$. As the spacetime is spherically symmetric, the geometry can be described by foliations of spherical 2-shells at any given instant. The Gaussian curvature of these spherical shells is
\begin{eqnarray}
K = - \E + \phi^{2}/4 \;.
\label{GaussCurv} 
\end{eqnarray}
From the above equations, it is clear that the electric part of the Weyl scalar is proportional to a $(3/2)th$ power of the Gaussian curvature, and the proportionality constant (which is the Schwarzschild mass `$m$') produces a covariant scale in the problem. We can also define the areal radius of the spherical 2-shells $r$, such that the Gaussian curvature is $1/r^2$. Integrating the propagation equations in terms of this variable, we get \cite{chris1}
\begin{eqnarray}
K=\frac{1}{r^2}~,~\phi=\frac{2}{r}\sqrt{F(r)}~,~\E=-\frac{2m}{r^3}~,~\A= - \frac{\E}{\phi} \, ,
\label{Schw1} 
\end{eqnarray}
where, $F(r)\equiv 1-2m/r$. These completely specify the pre-pulse geometry.

\subsection{The energy-momentum tensor of the EM test pulse}
\label{sec:EM-Tensor}

The test pulse contains the electromagnetic field that perturbs the background spacetime. In the semi tetrad 1+1+2 splitting, the electric and magnetic field vectors can be written in the following way:
\begin{eqnarray}
E^{a} & = & \Es e^{a} + \Es^{a}\;,\label{EField}\\
B^{a} & = & \Bs e^{a} + \Bs^{a}\;,
\end{eqnarray}
where $\Es$ and $\Bs$ are scalars and $\Es^{a}$ and $\Bs^{a}$ are projected 2-vectors on the 2-shells that break the sphericity of these shells. The energy-momentum tensor $T_{ab}$ of the field can then be split into a scalar, 2-vector, PSTF 2-tensor parts as
\begin{eqnarray}
T_{ab} = \mu u_{a}u_{b} + ph_{ab}+2 \Q \, e_{(a}u_{b)} +2 \Q_{(a}u_{b)}+\Pi(e_{a}e_{b}- \frac{1}{2}\N_{ab}) + 2\Pi_{(a}e_{b)} + \Pi_{ab}\;,\label{EMTensor}
\end{eqnarray}
where,
\ba 
\mu & = & \frac{1}{2} (\Es^2 + \Bs^2 + \Es^{a}\Es_{a} + \Bs^{a}\Bs_{a}) \;\\
p & = & \frac{1}{6} (\Es^2 + \Bs^2 + \Es^{a}\Es_{a} + \Bs^{a}\Bs_{a})\;\\
\Q & = & \veps_{ab}\, \Es^{b} \Bs^{c}\;\\
\Q_{a} & = & \veps_{ac}(\Bs \Es^{c} - \Es \Bs^{c}) \;\\
\Pi & = &  \frac{1}{3} \left[-2(\Es^2 + \Bs^2 ) + (\Es^{a}\Es_{a} + \Bs^{a}\Bs_{a})\right]\; \\
\Pi_{a} & = &  -\left(\Es \Es_{a} + \Bs \Bs_{a}\right)\; \\
\Pi_{ab} & = & \frac{1}{2} (\Es^{c}\Es_{c} + \Bs^{c}\Bs_{c}) \N_{ab} - (\Es_{a}\Es_{b} + \Bs_{a}\Bs_{b}) \;.
\ea
For the pulse not to back-react on the metric, the ratio of the magnitude of each component of the above stress-tensor w.r.t the background Gaussian curvature must be less than the BH mass ($m$). This has been verified in  ~Refs.~\cite{bir1,bir2,bir3}. The above energy-momentum tensor must follow the conservation equations up to the first-order perturbations on the background Schwarzschild manifold of the pre-pulse phase, which are given as
\ba 
\label{Energy}
\dot{\mu}+\hat{\Q}= & -\delta_a \Q^a-(\phi+2 \mathcal{A})\Q 
\ea
\ba
\label{Momentum1}
\dot{\Q} + \hat{p}+\hat{\Pi} =  \delta_a \Pi^a-\left(\frac{3}{2} \phi+\A \right) \Pi-(\rho + p) \A 
\ea
\ba
\label{Momentum2}
\dot{\Q}_{\bar{a}}+\hat{\Pi}_{\bar{a}}= & -\delta_a p+\frac{1}{2} \delta_a \Pi-\delta^b \Pi_{a b} -\left(\frac{3}{2} \phi+ \A \right) \Pi_a 
\ea
where the bar on the indices denotes the projected part on the perturbed 2-shells. 
\subsection{Pulse phase}
\label{sec:pulse-phase-eqs}

During this phase, the worldine $\gamma$ intersects the null cones of the test pulse for finite proper time. Due to the non-spherical pulse, the open sets around $\gamma$ will be non-spherical in the linear-order perturbations. Hence, as we show, it depends on the non-spherical energy-momentum tensor of the test pulse. In this phase, the spacetime is almost vacuum, spherical locally, and almost Schwarzschild. As shown in Refs.~\cite{bir1,bir2,bir3}, the rigidity of the vacuum spherically symmetric manifold in GR continues even in the perturbed scenario, and this almost Schwarzschild manifold will continue over the entire pulse phase.

{For the pulse not to back-react with the metric, the ratio of the magnitude of each component of the above stress-tensor w.r.t the background Gaussian curvature must be less than the BH mass ($M$). This has been verified in  ~Refs.~\cite{bir1,bir2,bir3}.}
Due to the pulse, the perturbed field equations contain both the background quantities and the first-order variables that determine the non-sphericity of the geometry. The total number of geometrical and  {EM pulse} variables are given by $\mathcal{D}\equiv\mathcal{D}_0\bigcup \mathcal{D}_1^{\rm geom}\bigcup\mathcal{D}_1^{\rm EM}$, where
\begin{eqnarray}
\mathcal{D}_1^{\rm geom}
& = &
\left\{\Theta, \sss\Omega, \sss\Sigma, \sss\mathcal{H}, \sss\xi, \sss\mathcal{A}_{a}, \sss\Omega_{a}, \sss\Sigma_{a}, 
\sss\alpha_{a}, \sss a_{a}, \sss\mathcal{E}_{a}, 
\sss\mathcal{H}_{a},\right. 
\left.\sss\Sigma_{ab}, \sss\zeta_{ab}, 
\sss\mathcal{E}_{ab}, \sss\mathcal{H}_{ab}\right\} \\
%
\mathcal{D}_1^{\rm EM}&=&\brac{\mu, \sss p, \sss \Q, \sss \Pi, \sss \Q_{a}, \sss \Pi_{a}, \sss \Pi_{ab}}.
\end{eqnarray}
Physical observables in GR must be gauge-invariant~\cite{1996-vanElst.Ellis-CQG}. We must identify the gauge-invariant variables from the above set to compare the derived quantities with observations. {Pulse introduces variations in $u^a$ and $n^a$ along spacelike and timelike directions, rendering the variables in the set $\mathcal{D}_0$ non-gauge-invariant.} As per the Stewart-Walker lemma, the quantities that vanish in the background spacetime are automatically gauge-invariant~\cite{SW,EllisBruni}.
For complete gauge invariance, we replace these variables with the following three~\cite{chris,anne}
\begin{eqnarray}
\mathcal{D}_1^{\rm GI}=\left\{W_a =\delta_a \E,  Y_a =\delta_a \phi, Z_a =\delta_a \A \right\}.
\end{eqnarray}
Hence, the complete set of first-order variables $\mathcal{D}_1^{\rm GI}\bigcup \mathcal{D}_1^{\rm geom}\bigcup\mathcal{D}_1^{\rm EM}$, determine the perturbed spacetime in a covariant and gauge-invariant way. For these variables, linearized evolution and propagation equations can be written as follows:
\ba 
\mathcal{L}^{(\rho)} \mathcal{D}^{\rm GI} = \mathcal{D}^{\rm geom} + f(\mathcal{D}^{\rm EM}) \, ,
\label{eq:generic1}
\ea
where, $\mathcal{L}^{(\rho)}$ represents evolution, propagation, or projected covariant derivatives on $2-$space. 
For details, see the Appendix.~\eqref{sec:CovariantFormalism}. One key feature of these equations is that the stress-tensor of the EM field (at the instant when the pulse reaches the worldline $\gamma$) sources these linearized equations. Since the $\mathcal{D}^{\rm geom}$ variables vanish before the pulse arrives, {\em the non-trivial initial data that generated these gauge-invariant variables in the pulse phase are completely supplied by the EM pulse.}

In the pulse phase, the first-order evolution equations for the above-defined gauge invariant first-order variables depicting the non-sphericity of the manifold can be written as follows. Here, the curly brackets denote the projected symmetric trace-free part of the tensor on the 2-sheet. For simplicity for the readers, we have put all the EM contributions within the square bracket of the RHS of each equation. This is to transparently show which terms will identically vanish at the following (post-pulse) phase despite the continuing non-sphericity.
\bea
\dot W_a &=&\frac32 \phi \,\E \bra{\alpha_a + \Sigma_a - \veps_{ab} \Omega^b}
+ \frac32 \E\bra{ \delta_a \Sigma - \frac23 \delta_a \Theta}+\, \veps_{bc} \delta_a \delta^b \H^c   \nn \\
&&-\left[\frac12 \delta_a \dot{\Pi} + \frac12 Y_{a} \Q + \frac12 \phi \left(\delta_{a} \Q \right) + \frac13 \delta_a \dot{\mu} + \delta_a\delta_b \Q^b\right]
\label{newEdot} ~, 
\eea
\bea
\dot Y_a &=& \bra{\frac12 \phi^2 + \E}\bra{\alpha_a + \Sigma_a - \veps_{ab} \Omega^b}
+ \delta_a\delta_c \alpha^c +\,\bra{\frac12 \phi - \A}\bra{\delta_a \Sigma - \frac23 \delta_a \Theta}+\left[\delta_a\Q\right]   ~.
 \label{newphidot} 
\eea

The shear evolution equations give
\bea
\label{eq:shearevol01}
\dot\Sigma_{a}-\f12\hat\udota_{a}&=&\f12\delta_a\udota+\bra{\udota-\f14\phi}\udota_a
    +\f12\udota a_a-\f32\Sigma\alpha_a-\E_a + \left[\frac{1}{2} \Pi_{a}\right]\;,
\eea
\bea
\label{eq:shearevol02}
\dot\Sigma_{\lb ab\rb}&=&\delta_{\lb a}\udota_{b\rb}
+\udota\zeta_{ab} -\E_{ab} + \left[\frac{1}{2} \Pi_{ab}\right]\;.
\eea

Evolution equation for $\hat e_a$ is:
\bea
\hat\alpha_{a}-\dot a_{a}&=&\bra{\f12\phi-\udota}\bra{\Sigma_a+\lc_{ab}\Omega^b}-\bra{\f12\phi+\udota}\alpha_a-\lc_{ab}\H^b + \left[\frac{1}{2} \Q_{a}\right].
\label{hatalphanl}
\eea

Electric Weyl evolution gives
\bea\label{replace2}
\dot\E =\bra{\f32\Sigma-\Theta}\E+\lc_{ab}\delta^a\H^c +\left[\frac{1}{3}\left(\frac{1}{2}\phi - 2\A \right)\Q - \frac12 \dot{\Pi} - \frac13\hat{\Q}+ \frac16 \delta_{a}\Q^a\right]
\eea
\bea
 \dot\E_{a}+\f12\lc_{ab}\hat\H^b &=&
    \f34\lc_{ab}\delta^b\H+\f12\lc_{bc}\delta^b\H^c_{~a}
    -\f34\E(\Sigma_a+2\alpha_a) 
    +\f34\E\lc_{ab}\Omega^b
     -\bra{\f14\phi+\udota}\lc_{ab}\H^b \nonumber\\&&-\left[\frac12 \dot{\Pi}_{a} + \frac14 \hat{\Q}_{a}+\frac{1}{4}\delta_{a} \Q -\frac{1}{2}(\frac{1}{4}\phi - \A) \Q_{a}\right]\;,
\eea
\bea
\dot\E_{\lb ab\rb}-\lc_{c\lb a}\hat\H_{b\rb}^{~~c} &=&
    -\lc_{c\lb a}\delta^c\H_{b\rb}
    -\f32\E\Sigma_{ab} 
    +\bra{\f12\phi+2\udota}\lc_{c\lb a}\H_{b\rb}^{~~c}  -\left[ \frac12 \dot{\Pi}_{\lb a b \rb}+\frac{1}{2} \delta_{\lb a}\Q_{b\rb}\right]\;.
\eea

Magnetic Weyl evolution gives
\bea
\dot\H&=&-\lc_{ab}\delta^a\E^b-3\xi\E +\left[\frac{1}{2}\lc_{ab}\delta^{a}\Pi^{b}\right]\;,
\eea
\bea
\dot\H_{a}-\f12\lc_{ab}\hat\E^b&=&-\f32\E\lc_{ab}\udota^b 
    +\f34\E\lc_{ab}a^b-\f12\lc_{bc}\delta^b\E^c_{~a} 
    +\bra{\f14\phi+\udota}\lc_{ab}\E^b
   -\f34\lc_{ab}\delta^b\E\nonumber\\&&
   -\left[ \frac{1}{4}\lc_{ab}\hat{\Pi}^{b}-\frac{3}{8}\lc_{ab}\delta^{b}\Pi - \frac{1}{4}\lc_{bc}\delta^{b}\Pi^{c}\,_{a}\right]\;,
\eea
\bea
\dot\H_{\lb ab\rb}+\lc_{c\lb a}\hat\E_{b\rb}^{~~c} &=&
    +\f32\E\lc_{c\lb a}\zeta_{b\rb}^{~~c}
    -\bra{\f12\phi+2\udota}\lc_{c\lb a}\E_{b\rb}^{~~c} 
     +\lc_{c\lb a}\delta^c\E_{b\rb}\nonumber\\&&
    + \left[\frac{1}{2}\lc_{c \lb a}\hat{\Pi}_{b\rb}\,^{c}- \frac{1}{2}\lc_{c \lb a}\delta^{c}\Pi_{b\rb}\right]
     \;.
\eea

The time evolution equations for 
$\xi$ and $\zeta_{\lb ab\rb}$  are 
\be 
\dot\xi =\bra{\udota-\f12\phi}\Omega
+\f12\lc_{ab}\delta^a\dotn^b+\f12\H\;, \label{dotxinl} 
\ee 
\bea
\dot\zeta_{\lb ab\rb}&=&\bra{\udota-\f12\phi}\Sigma_{ab} +\delta_{\lb
a}\alpha_{b\rb} -\lc_{c\lb a}\H_{b\rb}^{~~c}\;. \label{dotzetanl}
\eea

The vorticity evolution equations are  
\bea
\dot\Omega&=&\f12\lc_{ab}\delta^a\udota^b+\udota\xi \;,
\eea 
\bea
\dot\Omega_{a}+\f12\lc_{ab}\hat\udota^b&=&\f12\lc_{ab}\bra{-\udota
    a^b+\delta^b\udota-\f12\phi\udota^b}\;.
\eea

Sheet expansion evolution is given by:
\bea\label{replace3}
\dot\phi&=&\bra{\f23\Theta-\Sigma}\bra{\A-\f12\phi}+\delta_a\alpha^a\, + \left[\Q \right].
\eea

The Raychaudhuri equation is
\bea\label{replace4}
\hat\A-\dot\Theta=-\delta_a\A^a-\bra{\A+\phi}\A\, + \left[ \frac{1}{2}(\mu + 3p) \right].
\eea

The
propagation equations of $\xi$ and $\zeta_{\lb ab\rb}$ are: 
\bea
\hat\xi&=&-\phi \xi +\f12\lc_{ab}\delta^aa^b\;,
\eea 
\bea \hat\zeta_{\lb ab\rb}&=&-\phi\zeta_{ab} +\delta_{\lb
a}\hatn_{b\rb } -{\cal
E}_{ab} -\left[ \frac{1}{2}\Pi_{ab}\right]\;.
\label{hatzetanl} 
\eea

The shear divergence is given by : 
\bea\label{sigmahat}
\hat\Sigma-\f23\hat\Theta&=&-\f32\phi\Sigma-\delta_a\Sigma^a -\left[ \Q \right]
\eea
\bea
\hat\Sigma_{a}-\lc_{ab}\hat\Omega^b&=&\f12\delta_a\Sigma
    +\f23\delta_a\theta - \lc_{ab}\delta^b\Omega-\f32\phi\Sigma_a
     -\f32\Sigma a_a\nonumber\\&&
    +\bra{\f12\phi+2\udota}\lc_{ab}\Omega^b
    -\delta^b\Sigma_{ab} - \left[\Q_{a}\right]\;,
\eea
\bea
\hat\Sigma_{\lb ab\rb}&=&\delta_{\lb a}\Sigma_{b\rb} -\lc_{c\lb a}\delta^c\Omega_{b\rb}
    -\f12\phi\Sigma_{ab}\nonumber\\&&
    -\lc_{c\lb a}\H_{b\rb}^{~~c}\;.
\eea

The vorticity divergence equation is:
\bea
\hat\Omega&=&-\delta_a\Omega^a+\bra{\udota-\phi}\Omega\;.\label{hatOmSnl}
\eea

The Electric Weyl divergence is
\bea\label{replace5}
\hat\E  &=&-\delta_a\E^a-\f32\phi\E +\left[\frac{1}{3}\hat{\mu} - \frac{1}{2}\hat{\Pi} - \frac{1}{2}\delta_{a}\Pi^{a} - \frac{3}{4}\phi \Pi \right]\;,
\eea
\bea
\hat\E_{a} &=& \f12\delta_a\E -\delta^b\E_{ab}
      -\f32\E a_a-\f32\phi\E_a + \left[\frac{1}{3}\delta_{a}\mu  +\frac{1}{4} \delta_{a}\Pi - \frac{1}{2} \hat{\Pi}_{a} - \frac{1}{2}\delta^b\Pi_{ab} - \frac{3}{4}\phi \Pi_{a} \right]\;.
\eea

The Magnetic Weyl divergence is:
\bea
\hat\H&=& -\delta_a\H^a
    -\f32\phi\H-3\E\Omega - \left[\frac{1}{2}\lc_{ab}\delta^{a}\Q^{b}\right]\;,
\eea
\bea
\hat\H_{a} &=& \f12\delta_a\H-\delta^b\H_{ab}
    -\f32\E\lc_{ab}\Sigma^b
    +\f32\E\Omega_a\nonumber\\&&
    +\f32\Sigma\lc_{ab}\E^b
    -\f32\phi\H_a - \left[\frac{1}{2}\lc_{ab}\delta^b \Q - \frac14 \lc_{ab} \phi \Q^b - \frac{1}{2}\lc_{ab}\hat{\Q}^{b}\;\right].
\eea

The sheet expansion propagation is:
\bea\label{replace6}
\hat\phi&=&-\f12\phi^2+\delta_aa^a-\E -\left[\frac{2}{3} \mu +\frac{1}{2}\Pi\right]\;.
\eea

We also have the following constraints:
\bea
\delta_a\Omega^a+\lc_{ab}\delta^a\Sigma^b=\bra{2\A-\phi}\Omega+\H~,
\eea
\bea
\frac12\delta_a\phi-\veps_{ab}\delta^b\xi-\delta^b\zeta_{ab}&=&-\E_a -\left[\frac{1}{2}\Pi_{a} \right]~ ,
\label{divzetanl1}
\eea
\bea
\delta_a\Sigma-\frac23\delta_a\Theta+2\veps_{ab}\delta^b\Omega +2\delta^b\Sigma_{ab}&=&
    -\phi\bra{\Sigma_a-\veps_{ab}\Omega^b}\nn\\&& -2\veps_{ab}\H^b - \left[\Q_{a}\right]. 
    \label{divSigmanl1}
\eea

\section{Comparing results with 
Regge-Wheeler-Zerilli BH perturbation theory}
\label{sec:RWZ-comparison}
In this section, we show explicitly how the characteristics of the GW, such as frequency, and luminosity or the total GW energy produced due to interaction among a very highly energized EM wave and a Schwarzschild BH, match with the results obtained from Regge-Wheeler-Zerilli(RWZ) BH perturbation theory~\cite{1957-Regge.Wheeler-PRD,1970-Zerilli-PRD}. As per RWZ perturbation theory, if an object of mass $m$ falls into a spherically symmetric non-rotating BH of mass $M$ ($M > m $), then the perturbed BH of mass $M$ will radiate GW as a form of \emph{quasi-normal mode} (QNM). The following list contains the comparison between our result and the results produced from RWZ analysis.   
\begin{enumerate}
%
\item \emph{Frequency:} We obtain the relation for the angular frequency of the produced GW (in geometrized unit): 
\begin{eqnarray}
\label{eq:omega-geometrised}
\omega = \frac{\sqrt{3}}{\rs}; ~~=>
M\omega = \frac{1}{2\sqrt{3}}
\end{eqnarray}
This expression for $\omega$ matches with GW frequency obtained from RWZ analysis for $l =2$ mode. The corresponding frequency $f_{\rm GW}$ in non-geometrized unit following the conversion mentioned in~Ref.\cite{1999-Kokkotas-LivRev} we obtain:
\begin{eqnarray}
\label{eq:omega-non-geometrised}
f_{\rm GW} = \frac{1}{2\sqrt{3}} \, 2\pi\times 5142 \times \frac{M_{\odot}}{M} {\rm Hz}= 9.327 \left(\frac{M_{\odot}}{M}\right) {\rm kHz}
\end{eqnarray}
As per this relation for a $10 M_{\odot}$ BH the frequency of the fundamental GW mode will be $f_{\rm GW} = 0.9$ kHz. This result matches with Ref.~\cite{1999-Kokkotas-LivRev}.
\item \emph{Energy radiated:}
As per our analysis we obtain the total luminosity associated with the emitted GW will be, $\mathcal{L}_{\rm GW} = 
\frac{1}{100\pi^2} \frac{E_{\gamma}^2}{\mathcal{L}_0}\left(\frac{c }{\rs}\right)^2
$ Substituting $\rs = 2 G M/c^2$ and value of $\mathcal{L}_0$ we obtain,
\begin{eqnarray}
\label{eq:GW-luminosity}
\mathcal{L}_{\rm GW} = 1.13 \times 10^{14}~\left(\frac{E_{\gamma}}{M}\right)^2 {\rm erg/s}.
\end{eqnarray}
Now from this GW luminosity we can obtain the total amount of energy emitted as GW using the relation $\mathcal{L}_{GW} = \frac{\Delta E_{\rm GW}}{2 \Delta t_{\rm GW}}$, where $\Delta E_{\rm GW},~\Delta t_{\rm GW}$ are energy and duration of the GW emission respectively. The $\Delta t_{\rm GW} \sim 1/f_{\rm GW}= 1.075 \times 10^{-4} \left(\frac{M}{M_{\odot}}\right)$. Employing this expression we obtain,
\begin{eqnarray}
\label{eq:GW-energy1}
\Delta E_{\rm GW} = \mathcal{L}_{\rm GW} \times \left(2\Delta t_{\rm GW}\right) =  \frac{1.22  ~E_{\gamma}^2 }{M}\times 10^{-23}
\end{eqnarray}
As, the EM energy 
can be interpreted as $E_{\gamma}  = m_{\rm equiv} c^2$, we can write the above equation as, 

\begin{eqnarray}
\nn
\Delta E_{\rm GW} & =  \frac{1.22  ~ m_{\rm equiv}^2 c^4}{M} \times 10^{-23} \\
\label{eq:GW-energy2}
 & = 1.09 \times 10^{-3} \frac{m_{\rm equiv}^2 }{M} c^2.
\end{eqnarray}
This equation shows that our result(total GW energy emitted as GW from a perturbed black hole of mass $M$) matches with the result produced as per RWZ BH perturbation analysis~\cite{1992-Lee.S.Finn-PRD}.

\end{enumerate}

\bibliography{Final}

\end{document}